\documentclass[conference]{IEEEtran}
\usepackage{cite}
\usepackage{graphicx}
\usepackage{caption}
\usepackage{subcaption}
\usepackage{verbatim}
\usepackage{pgfplots}
\usepackage{pgf-umlsd}
\usepackage{amsmath}
\usepackage{algorithm}
\usepackage{algorithmic}

\pgfplotsset{compat=1.14}
\pgfplotsset{colormap/viridis}

\ifCLASSINFOpdf
\else
\fi
\begin{document}
%
\title{Garou: An Efficient and Secure Off-Blockchain Multi-Party Payment Hub}

\author{\IEEEauthorblockN{Yongjie Ye}
\IEEEauthorblockA{School of Data and\\Computer Science\\
Sun Yat-Sen University\\
Guangzhou, China\\
Email: yeyyj6@mail2.sysu.edu.cn}
\and
\IEEEauthorblockN{Weigang Wu}
\IEEEauthorblockA{School of Data and\\Computer Science\\
Sun Yat-Sen University\\
Guangzhou, China\\
Email: wuweig@mail.sysu.edu.cn}
}


\maketitle

\begin{abstract}
  To mitigate the scalability problem of decentralized cryptocurrencies such as Bitcoin and Ethereum, the payment channel, which allows two parties to perform secure coin transfers without involving the blockchain, has been proposed.
  The payment channel increases the transaction throughput of two parties to a level that is only limited by their network bandwidth. 
  Recent proposals focus on extending the two-party payment channel to the N-party payment hub. 
  Unfortunately, none of them can achieve efficiency, flexibility in the absence of a trusted third-party.
  
  In this paper, we propose Garou, a secure N-party payment hub that allows multiple parties to perform secure off-chain coin transfers.
  Except in the case of disputes, participants within the payment hub can make concurrent and direct coin transfers with each other without the involvement of the blockchain or any third-party intermediaries.
  This allows Garou to achieve both high-performance and flexibility.
  Garou also guarantees that an honest party always maintains its balance security against strong adversarial capabilities.
  To demonstrate the feasibility of the Garou protocol, we develop a proof of concept prototype for the Ethereum network.
  Our evaluation results show that the maximum transaction throughput of Garou is $20 \times$ higher than that of state-of-art payment hubs.
\end{abstract}


%

\section{Introduction}

Since the advent of Bitcoin \cite{nakamoto2008bitcoin} in 2008, decentralized cryptocurrencies have gained great popularity over the last ten years.
The combination of consensus algorithm (e.g., POW \cite{dwork1992pricing} and POS \cite{king2012ppcoin}) and tamper-resistant hash-linked chain of blocks, i.e., blockchain, allows secure coin transfers between mutually untrusting peers across the world.
Inspired by Bitcoin, other cryptocurrencies supporting the execution of Turing-complete smart contracts have emerged.
These cryptocurrencies allow coins to be managed by user-defined smart contracts of arbitrary complexity.
Among them, the most prominent one is Ethereum \cite{wood2014ethereum},
which uses Solidity as the programming language for its smart contracts.

However, the major inherent limitation of these cryptocurrencies lies in their performance.
For example, Bitcoin can only support up to 6$\sim$7 transactions per second while Ethereum supports up to 20 transactions per second \cite{croman2016scaling}.
The global nature of consensus algorithms makes it impossible to grain a magnitude growth of throughput with simple re-parameterization \cite{gervais2016security}.
Such a low transaction throughput is far from enough to support the widespread use of decentralized cryptocurrencies, let alone the execution of complex smart contracts.
In contrast, Visa processes up to 47,000 transactions per second \cite{trillo2013stress}.

To mitigate the scalability problem of decentralized cryptocurrencies, the payment channel \cite{decker2015fast, poon2016bitcoin}, which allows two parties to perform secure off-chain micropayments, has been proposed.
Except for the initialization and the closing process, all transactions between the two parties of a payment channel can be executed privately without committing to the blockchain.
Therefore, the payment channel can significantly increase the transaction throughput of two parties and reduce the storage burden for the blockchain system.
One extension of the two-party payment channel is the payment network, which uses the payment channels as building blocks to construct a linked off-chain overlay.
The payment network uses the \textit{Hashed Time-Lock Contract} (HTLC) transactions to route payments across multiple intermediary payment channels.
HTLC transactions allow coins to be transferred between parties that are not directly connected by a payment channel.
However, HTLC transactions require that every intermediary payment channel to lock a portion of its available capacity until the payment is settled, which may lead to deadlock in a concurrent situation \cite{malavolta2017concurrency}.

Recently efforts have been made to extend the two-party payment channel to N-party payment hub, which allows multiple parties to perform secure off-chain coin transfers.
TumbleBit \cite{ethan2017tumblebit} is a Bitcoin-compatible anonymous payment hub, which guarantees that no one could link a payment from its payer to its payee.
The reliance on the intermediary \textit{Tumbler}, however, may reduce its transaction flexibility and increase the risk of a single point of failure.
Dziembowski et al. \cite{dziembowski2019multi} propose a novel approach to recursively construct multi-party virtual state channel (which can be viewed as hub as well) on top of two-party state channels.
However, any state transition of the multi-party virtual state channel should be explicitly agreed upon by all participants, which significantly limits its performance in the payment scenario.
NOCUST \cite{khalil2018nocust} is a new specification for the N-party payment hub with a high degree of flexibility, which allows its participants to transfer all their allocated funds to any other member.
However, the execution of NOCUST also relies on a third-party operator server, which could potentially become a single point of failure.
Moreover, the requirement that the operator server should commit data to the blockchain periodically also de-incentivizes its election and may cause substantial overhead to the payment hub.

In summary, extending the two-party payment channel to the N-party payment hub is a nontrivial task.
None of the existing proposals can achieve both efficiency, flexibility, and untrusty.

\subsubsection*{This work}

In this paper, we propose a high-performance, flexible and untrusted N-party payment hub called Garou.
It allows multiple parties to make concurrent and flexible off-chain coin transfers with each other without compromising security.
Specifically, We adopt an automatic leader election mechanism to eliminates the reliance on any trusted third-party.
The separation of spending and receiving allows Garou to achieve high concurrency.
Besides, the escrow-free design allows participants to spend their money flexibly.
Finally, Garou allows its participants to join in and withdraw from the payment hub at any time. 

The main features of Garou are listed as follows:

\begin{itemize}
    \item \textbf{Efficiency.}
    Garou is designed for high-performance off-chain coin transfers.
    Concurrent transactions are allowed among participants of the payment hub.
    Except in the case of disputes, the off-chain transaction of Garou does not require the involvement of the blockchain, and therefore Garou can significantly increase the performance and scalability of the blockchain systems.
    Our evaluation results show that the maximum transaction throughput of Garou is $20 \times$ higher than that of state-of-art payment hubs.
    \item \textbf{Flexibility.}
    Garou exhibits a high level of transaction flexibility, which allows its participants to freely transfer all their available funds to any other member of the payment hub.
    It does not require any intermediary nodes to route coin transfers or escrow their coins before transactions.
    Besides, Garou allows its participants to join in or withdraw from the payment hub at any time.
    \item \textbf{Security.}
    Garou guarantees that an honest party will not bear any financial losses despite strong adversarial capabilities.
    Using the on-chain smart contract as a dispute resolver, an honest party can always maintain its balance security even when all other members of the payment hub are malicious.
    Besides, Garou does not rely on any third-party operators and therefore can effectively avoid the single points of failure.
\end{itemize}


\subsubsection*{Organization of the paper}

In Section \ref{background}, we provide the necessary background and review the state-of-art payment hub designs.
In Section \ref{garouProtocol}, we present an overview and the detailed design of the Garou protocol.
The security of Garou is analyzed in Section \ref{securityAnalysis}.
In Section \ref{implementation}, we provide our proof-of-concept implementation and the evaluation results based on the Ethereum network.
Finally, we conclude this paper in Section \ref{conclusion}.

\section{Background and Related Works} \label{background}

Decentralized cryptocurrencies such as Bitcoin \cite{nakamoto2008bitcoin} and Ethereum \cite{wood2014ethereum} have been suffering from severe performance and scalability problems since their birth.
Recently many attempts have been made to mitigate these problems such as
alternative consensus mechanisms \cite{eyal2016bitcoin, luu2015scp, pass2017hybrid},
usage of trusted execution environment \cite{lind2017teechain, zhang2016town},
blockchain sharding \cite{zamani2018rapidchain, luu2016secure, kokoris-kogias2018omniledger, nguyen2019optchain},
payment channel and payment network \cite{decker2015fast, poon2016bitcoin, mccorry2016towards}, etc.

The sharding technique splits the blockchain into multiple shards, and each shard forms a consensus group and maintains its fraction of nodes or transactions.
This could improve the performance of the blockchain system to thousands of transactions per second and reduces confirmation time to a few seconds \cite{nguyen2019optchain}.
One advantage of the blockchain sharding is that it does not sacrifice the transaction traceability, i.e., all transactions still need to be committed to the blockchain.
Blockchain sharding, however, also faces many challenges, such as cross-shard atomicity \cite{wang2019monoxide}, transaction placement \cite{nguyen2019optchain}, single-shard takeover attack \cite{chauhan2018blockchain}, etc.

Another approach to improving the scalability of cryptocurrencies is the off-chain payment channel/network.
At the cost of transaction traceability, the off-chain techniques can achieve substantial performance improvement and lower storage burden than the sharding techniques.

\subsection{Payment Channels}

Payment channels \cite{decker2015fast, poon2016bitcoin, decker2018eltoo} allow two parties to perform secure coin transfers privately without involving the blockchain, yet still maintaining balance security of honest parties.
The life cycle of a payment channel is divided into three stages: opening, trading, and closing.
The opening stage requires the two parties of the payment channel to commit a funding transaction to the blockchain, together with their deposits on the payment channel.
After the payment channel is successfully opened, two parties can perform off-chain transactions privately without committing them to the blockchain.
At any point, each of the two parties can commit the final state of trading to the blockchain to close the payment channel and reclaim their share of deposits.
Because all coin transfers of the payment channel are performed privately and not committed to the blockchain, the payment channel can significantly increase the transaction throughput of two parties to a level that only limited by their network bandwidth, as well as reducing the storage burden for the blockchain system.
The key challenge of payment channel protocol is to enforce the latest off-chain transaction state in the presence of malicious parties.
Decker et al. \cite{decker2015fast} use the blockchain-based \textit{time locks} to invalidate obsolete transaction state.
The Lightning Network \cite{poon2016bitcoin} resorts to punishment to prevent malicious nodes from rolling back.

There are several improvement proposals on payment channel protocol.
Green et al. \cite{green2017bolt} propose \textit{Blind Off-chain Light-weight Transactions} (Bolt) to construct privacy-preserving and unlinkable payment channels for anonymous cryptocurrencies such as ZCash \cite{hopwood2016zcash, miers2013zerocoin}.
Perun \cite{dziembowski2019perun} is a virtual payment channel built on top of existing ledger channels to enabling efficient off-chain transactions between two parties that are connected to one intermediary Ingrid.
The state channel \cite{dziembowski2018general} proposes a novel technique to recursively construct a virtual state channel that allows off-chain execution of arbitrary complex smart contracts.

\subsection{Payment Networks}

Using the two-party payment channel as building blocks, we could construct a linked payment network \cite{poon2016bitcoin, mccorry2016towards}, where nodes can utilize multiple intermediary payment channels to establish a multi-hop path and route coins to others.
One of the key challenges of the payment network is to enforce atomicity when routing coins through multiple payment channels.
To achieve this, the Lightning Network \cite{poon2016bitcoin} adopts a technique called \textit{Hashed Time-Lock Contract} (HTLC) transaction.
An HTLC transaction is an off-chain conditional contract that locks a portion of coins of a payment channel.
These locked coins can be reclaimed by the receiver once it provides the qualified pre-image, or they are returned to the owner after times out.

In addition to atomicity, some other factors have also been concerned, such as privacy, concurrency, and routing efficiency.
Malavolta et al. \cite{malavolta2019anonymous} proposed a novel cryptographic primitive called \textit{Anonymous Multi-Hop Locks (AMHLs)} to ensure privacy when routing coins through the payment networks.
They also provide an in-depth discussion about the privacy and concurrency in payment networks and design the first non-blocking solution \cite{malavolta2017concurrency}.
Flare \cite{prihodko2016flare}, SilentWhispers \cite{malavolta2017silentwhispers}, and SpeedyMurmurs \cite{roos2018settling} mainly focus on improving efficiency and privacy of the routing process in the payment networks.

\subsection{Payment Hubs}

In this paper, the term ``payment hub'' refers to an infrastructure that allows multiple parties to make off-chain coin transfers with each other.
The multi-party payment hub can be regarded as an extension of the two-party payment channel.
In the following, we review existing works on the payment hub and compare them with our proposed one.

\subsubsection{TumbleBit}

TumbleBit \cite{ethan2017tumblebit} is an anonymous payment hub designed for the Bitcoin network.
It uses cryptographic techniques to ensure the privacy and anonymity of its participants, i.e., no one could link a payment from its payer to its payee.
TumbleBit requires that each participant open a directed payment channel with the untrusted intermediary, which is called Tumbler.
To make a coin transfer, the payer and the Tumbler should first escrow some coins on their payment channel, and then the payer and payee run a puzzle-solver protocol and exchange cryptographic materials to reclaim their share of escrowed coins while preventing the Tumbler from violating their anonymity.

The reliance on the Tumbler, however, could lead to a dangerous situation where a malicious Tumbler could fully control the transactions of the payment hub.
Besides, the requirement that the Tumbler should escrow the same amounts of coins before transfers also undermines the flexibility of the payment hub.
This prevents members from freely spending their balance when the Tumbler does not have enough deposit to be escrowed in its payment channels.
In contrast, our proposed Garou protocol exhibits a high level of flexibility, which allows the available funds of a participant to be freely spent to any other member of the payment hub.
Garou does not require any intermediary nodes to route coin transfers or escrow their coins before transactions.

\subsubsection{Multi-party Virtual State Channel}

Dziembowski et al. \cite{dziembowski2019multi} propose a novel approach to recursively construct multi-party virtual state channel (which can be viewed as hub as well) on top of two-party state channels.
Once the underlying state channels are established, the upper layer of virtual state channels can be created and closed at negligible costs.
The state channel allows the off-chain execution of arbitrary complex smart contracts, which significantly enrich the functionality of the off-chain ecosystem.
However, any state transition of the multi-party virtual state channel must be explicitly agreed by all participants, which leads to poor performance in the payment scenario.
In contrast, our proposed payment hub is dedicated to off-chain coin transfers.
Each transfer only requires the involvement of three participants and the off-chain transactions are confirmed in batch mode.
This can significantly reduce consensus costs and increase the transaction throughput of the payment hub.

\subsubsection{NOCUST}

NOCUST \cite{khalil2018nocust} is an N-party payment hub with a high degree of flexibility, which allows the allocated funds of a party to be totally paid to any other member of the payment hub.
A NOCUST payment hub consists of two fundamental components: an on-chain verifier contract and an off-chain operator server.
The on-chain verifier contract serves as a trusted financial custodian which manages the deposits of all participants and acts as a dispute resolver.
The off-chain operator server executes every coin transfer and periodically commits the balance information of all participants (encoded into a merklelized interval tree) to the on-chain verifier contract to keep consistency.

The major drawback is that the execution of the NOCUST payment hub relies on a third-party operator server, exposing the system to the single point of failure.
The operator server is also required to make costly on-chain data commitments periodically, which de-incentivizes its election, causes substantial overhead to the payment hub, and significantly limits the scalability of the blockchain system.
Moreover, NOCUST does not allow its participants to initiate any other transfers until the current one is done, which limits the concurrency of the payment hub.
In contrast, our proposed protocol does not rely on any third-party operators.
All operations can be done internally.
Besides, Garou allows concurrent coin transfers between participants and therefore Garou can achieve high-performance.

\section{Garou Protocol} \label{garouProtocol}

In this section, we firstly present the system model and then describe the design of the Garou protocol.
Garou is dedicated to off-chain coin transfers and provides two basic functionalities: 
(1) dynamic enrollment and withdrawal; 
(2) secure and concurrent off-chain coin transfers;

\subsection{System Model} \label{systemModel}

\subsubsection{Blockchain and Smart Contract} \label{blockchain}

In this paper, we model the blockchain as a trusted and immutable ledger that records all valid on-chain transactions.
Those transactions will be globally available for all participants after being confirmed on the blockchain, usually on an average block time \textit{T}.
In a typical blockchain network, every participant is uniquely identified by its account address, which is usually hash-derived from a public key.

In addition to regular coin transfers between participants, many blockchain systems also support the execution of arbitrary complex smart contracts.
Our proposed protocol requires a smart contract execution environment such as Ethereum \cite{buterin2014ethereum}.
Smart contracts allow money to be fully controlled by its program code.
Once a smart contract is deployed on the blockchain, its public functions can be invoked by anyone via an on-chain transaction, which is processed by the miners.
Transaction fees have to be paid to incentivize miners to process the transaction and execute the smart contracts.
Besides, the same piece of smart contract code can be deployed multiple times. 
Each contract instance has its own globally unique identifier (known as \textit{contract account} in the Ethereum network), and different contract instances are independent of each other.

\subsubsection{Network Model} \label{communicationModel}

Given that every participant in the blockchain network is uniquely identified, we assume an underlying communication network that all participants of the payment hub can communicate directly with each other.
The communication between participants is authenticated and confidential, e.g. through TLS.
We further assume that the communication channels between participants are synchronous, i.e., messages sent between honest participants arrive within a maximum delay $\Delta$.
Finally, we assume that the connections between any two uncompromised participants and the communication with the blockchain network are uncorrupted.
We note that all these above assumptions are reasonable and held in almost all blockchain systems.

\begin{table}[!t]
    \renewcommand{\arraystretch}{1.3}
    \caption{Information Maintained by Leader}
    \label{tbl:notaion}
    \centering
    \begin{tabular}{|c||l|}
    \hline
    \bfseries Notation & \bfseries Description \\
    \hline\hline
      $\mathcal{B}_{i}(e)$ & Initial balance of $\mathcal{P}_i$ of epoch $e$ \\
      \hline
      $\mathcal{S}_{i}(e)$ & Total amount successfully sent by $\mathcal{P}_i$ during epoch $e$ \\
      \hline
      $\tau_{i}(e)$ & Total amount tried to be sent by $\mathcal{P}_i$ during epoch $e$ \\
      \hline
      $\mathcal{R}_{i}(e)$ & Total amount received by $\mathcal{P}_i$ during epoch $e$ \\
      \hline
      $\mathcal{E}(e)$ & Enrollment set of epoch $e$ \\
      \hline
      $\mathcal{W}(e)$ & Withdrawal set of epoch $e$ \\
    \hline
\end{tabular}
\end{table}

\subsection{Overview} \label{overview}

We now describe the key ideas of Garou and give a high-level overview on how Garou achieve both high-performance, flexibility, untrusty, and security.

Garou is composed of two fundamental building blocks: an on-chain smart contract and an off-chain transaction protocol.
The on-chain smart contract manages the deposits of all participants of the payment hub and acts as a dispute resolver.
It guarantees that the latest epoch state agreed by all participants is always accepted.
Backed by the on-chain smart contract, an honest party will not bear any financial losses even when all other participants of the payment hub are malicious.

The off-chain transaction protocol is designed for concurrent and flexible off-chain coin transfers and consists of multiple epochs, each of which is divided into three phases: the leader election, trading, and consensus phase.
(1) For each epoch, a leader is elected out of all participants to drive the off-chain execution of the protocol.
We adopt an automatic leader election mechanism to eliminate additional communication overhead while avoiding the single point of failure caused by a fixed leader.
(2) During the trading phase, participants of the payment hub can freely make coin transfers with each other.
Garou does not require any intermediary nodes to route coin transfers or escrow their coins before transactions, and therefore Garou exhibits a high degree of flexibility.
Moreover, the spending and receiving of coins are separated, i.e., participants cannot spend more than their initial balance of the current epoch, and its receiving coins cannot be spent until the next epoch.
This simplifies the verification and allows participants to perform concurrent transactions.
(3) Finally, during the consensus phase, all participants collaborate to reach a consensus about the final state of the current epoch, which acts as a checkpoint for the off-chain execution of the Garou protocol.
In case of disputes, an honest participant could commit the latest epoch state to the on-chain smart contract to enforce its balance security.

\subsection{Detailed Description} \label{detailedDescription}

In this paper, we denote the set of participants as $\mathcal{P} = \{\mathcal{P}_1, \ldots, \mathcal{P}_n\}$, whereby all participants are ordered according to the time of joining.

\subsubsection{Leader Election Phase} \label{ledaerElectionPhase}

At the beginning of each epoch, a leader needs to be elected out of all participants to drive the off-chain execution.
Specifically, the leader is responsible for: 
\begin{itemize}
	\item maintaining information as shown in Table \ref{tbl:notaion};
	\item issuing transaction id for every off-chain transaction in the current epoch;
	\item early verification of every transaction to prevent malicious participant from double-spending;
	\item driving all participants of the payment hub to collaborate to reach a consensus on the epoch state during the consensus phase;
	\item answering the epoch state challenge in case of disputes (discussed in Section \ref{onChainDisputeResolver}).
\end{itemize}
To avoid the single point of failures caused by a fixed and malicious leader, we adopt a leader election mechanism to automatically elect a leader according to the balance information of the current epoch.
More formally, the leader of epoch $e$ is elected by:
\begin{equation}
 \label{eqn:leaderElection}
 l = H(\bigoplus^n_{i=1} \mathcal{B}_i(e)) \mod n
\end{equation}
where $H$ is an arbitrary hash function, $\bigoplus$ denotes the XOR operation, $\mathcal{B}_i(e)$ denotes the initial balance of participant $\mathcal{P}_i$ in epoch $e$ (as shown in Table \ref{tbl:notaion}).
By this way, participant $\mathcal{P}_l$ will automatically become the leader of epoch $e$.

Garou requires that all participants should maintain knowledge about the initial balance $\mathcal{B}(e)$ for every epoch, and therefore they can determine the leader by themselves without any additional communication or synchronization overhead, i.e., the leader election can be done at negligible cost.

\subsubsection{Trading Phase} \label{tradingPhase}

\begin{figure}[!t]
 \centering
    \begin{tikzpicture}
    [->,>=stealth', shorten >=1pt, auto, thick,
      node distance=4.5cm,
      main node/.style={circle, fill=blue!20, draw, minimum size=10mm}
    ]
    \node[main node] (L) {\bfseries{$\mathcal{P}_{l}$}};
    \node[main node] (S) [below left of=L] {\bfseries{$\mathcal{P}_i$}};
    \node[main node] (R) [below right of=L] {\bfseries{$\mathcal{P}_j$}};
    \path[every node/.style={
        font=\sffamily\small,
        fill=white, 
      inner sep=1pt}]
    (S) edge [bend left=20] node[anchor=center, pos=0.4] {(1) txIdReq} (L)
        edge [bend left=20] node[anchor=center, pos=0.5] {(3) txReq} (R)
        edge [bend left=60] node[pos=0.5] {(5) tx} (L)
    (L) edge [bend left=20] node[pos=0.2] {(2) txIdReply} (S)
    (R) edge [bend left=20] node[anchor=center, pos=0.5] {(4) txReply} (S);
  \end{tikzpicture}
    \caption{Message Flow of an Off-Chain Transaction. $\mathcal{P}_l$ denotes the leader of the current epoch; $\mathcal{P}_i$ denotes the sender and $\mathcal{P}_j$ is the receiver of the transaction.}
    \label{fig:trading}
\end{figure}
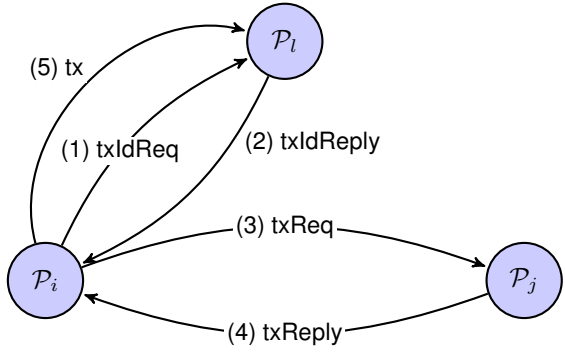

During the trading phase, participants within the payment hub are free to make coin transfers with each other.
We use the following tuple to denote an off-chain coin transfer:
\begin{equation}
 \label{eqn:tx}
 \textit{tx} = (e, \textit{txId}, \textit{sender}, \textit{recevier}, x)
\end{equation}
where \textit{txId} denotes the transaction id issued by the leader of epoch $e$, $x$ denotes the amount of coins to be transferred.
The transaction id is used for ordering transactions in the consensus phase.

As outlined in Fig. \ref{fig:trading}, to start an off-chain coin transfer, the sender $\mathcal{P}_i$ first requests a transaction id from the leader.
When the leader receives the request for a new transaction id, it should check if the sender is trying to over-spend its balance.
More formally, the leader should ensure the following holds:
\begin{equation}
 \label{eqn:leaderTxIdVerification}
 x \leq B_i(e) - \tau_i(e)
\end{equation}
If this is the case, the leader should approve this transaction, update the amount of coins that tried to be sent by $\mathcal{P}_i$ so that 
$\tau_i^{\prime}(e) = \tau_i(e) + x$, 
and then reply with a monotonically increasing transaction id signed by the leader itself.
After that, $\mathcal{P}_i$ sends a transaction request to the receiver $\mathcal{P}_j$, asking for his approval.
When receiving the transaction request, $\mathcal{P}_j$ should verify the leader's signatures on the transaction id and then reply to $\mathcal{P}_i$ with its own signatures on the transaction request.
Finally, $\mathcal{P}_i$ forwards the completed transaction request to the leader.
And the leader will update its information table so that
$\mathcal{S}_i^\prime(e) = \mathcal{S}_i(e) + x$ and
$\mathcal{R}_j^\prime(e) = \mathcal{R}_j(e) + x$.

Note that the off-chain transaction does not require any intermediaries to route or escrow their coins before transactions.
The sender $\mathcal{P}_i$ could spend all its available balance according to its own will.
From this perspective, Garou exhibits a high level of flexibility.
From the equation (\ref{eqn:leaderTxIdVerification}), we also notice that when doing the early verification of a transaction, the leader does not take into account the amount of coins received by $\mathcal{P}_i$ in the epoch $e$, i.e. $\mathcal{R}_{i}(e)$.
That is because of the separation of spending and receiving.
Since an off-chain transaction cannot be completely confirmed before the consensus phase, spending those unconfirmed coins immediately in current epoch would increase the risk of double-spending. 
Therefore, Garou does not allow coins received in the current epoch to be spent until the next epoch.
This limitation significantly simplifies the verification and enables us to perform concurrent transactions.
As a result, $\mathcal{P}_i$ can initiate another transaction before he sends the result of the current transaction to the leader, as long as he does not try to spend more than his initial balance of the current epoch.

\subsubsection{Consensus Phase} \label{consensusPhase}

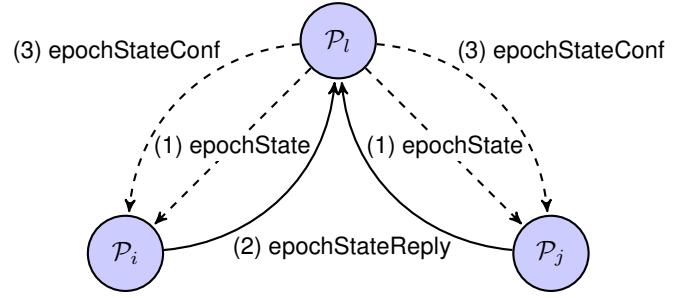
\begin{figure}[!t]
 \centering
    \begin{tikzpicture}
    [->,>=stealth', shorten >=1pt, auto, thick,
      node distance=4cm,
      main node/.style={circle, fill=blue!20, draw, minimum size=10mm}
    ]
    \node[main node] (L) {\bfseries{$\mathcal{P}_{l}$}};
    \node[main node] (S) [below left of=L] {\bfseries{$\mathcal{P}_i$}};
    \node[main node] (R) [below right of=L] {\bfseries{$\mathcal{P}_j$}};
    \path[every node/.style={
        font=\sffamily\small,
        fill=white,
      inner sep=1pt}]
    (L) edge [dashed] node[anchor=center, pos=0.5] {(1) epochState} (S)
    edge [dashed] node[anchor=center, pos=0.5] {(1) epochState} (R)
    edge [bend right=40, dashed] node[anchor=center, pos=0.3, above left] {(3) epochStateConf} (S)
    edge [bend left=40, dashed] node[pos=0.3] {(3) epochStateConf} (R)
    (S) edge [bend right=40] node[pos=0.25, below right] {(2) epochStateReply} (L)
    (R) edge [bend left=40] node {} (L);
  \end{tikzpicture}
    \caption{Communication of the Consensus Phase. The solid line indicates unicast and the dashed line indicates broadcast}
    \label{fig:consensus}
\end{figure}

In the consensus phase, all participants should collaborate to reach a consensus about the final state of the current epoch, aka the initial state of the next epoch, denoted as $\textit{State}(e+1)$.

The epoch state $\textit{State}(e)$ contains the following information:
\begin{equation}
 \label{eqn:state}
 \textit{State}(e) = (e, \mathcal{B}(e), \mathcal{M}(e-1), \mathcal{E}(e), \mathcal{W}(e))
\end{equation}
where $\mathcal{B}(e)$ denotes the balance of all participants of the payment hub and is calculated by:
\begin{equation}
 \label{eqn:balance}
 \mathcal{B}_{i}(e) = \mathcal{B}_{i}(e-1) - \mathcal{S}_{i}(e-1) + \mathcal{R}_{i}(e-1), i \in \{1, \ldots, n\}
\end{equation}
The transaction merkle tree root $\mathcal{M}(e)$ \cite{merkle1987digital} is constructed for each participant $\mathcal{P}_i$ using all its involved transactions in epoch $e$, including both the spending and receiving transactions (c.f. Algorithm \ref{alg:txMerkleTree}).
   
\begin{algorithm}[!t]
	\caption{Transaction Merkle Tree \textit{TMT}}
	\label{alg:txMerkleTree}
	\begin{algorithmic}[1]
		\renewcommand{\algorithmicrequire}{\textbf{Input:}}
		\renewcommand{\algorithmicensure}{\textbf{Output:}}
		\REQUIRE transaction set $(\textit{tx}_1, \ldots, \textit{tx}_n)$, ordered by transaction id
		\ENSURE transaction merkle tree root R
		\IF {$n = 1$}
		\RETURN $H(\textit{tx}_1)$
		\ELSE
		\STATE $\textit{left} \leftarrow \textit{TMT}(\textit{tx}_1, \ldots, \textit{tx}_{\lceil n/2 \rceil})$
		\STATE $\textit{right} \leftarrow \textit{TMT}(\textit{tx}_{\lceil n/2 \rceil + 1}, \ldots, \textit{tx}_{n})$
		\RETURN $H(\textit{left} \bigoplus \textit{right})$
		\ENDIF
	\end{algorithmic}
\end{algorithm}

As shown in Fig. \ref{fig:consensus}, the consensus phase is driven by the leader, who calculates the epoch state $\textit{State}(e+1)$, collects participants' signatures on $\textit{State}(e+1)$, and finally broadcasts the epoch state confirmation to all participants.
When participant $\mathcal{P}_i$ receives the epoch state from the leader, it should verify its own balance $\mathcal{B}_{i}(e+1)$, the proper construction of its transaction merkle tree root $\mathcal{M}_{i}(e)$, and ensure that $\sum_{k=1}^{n}\mathcal{B}_{k}(e+1)$ equals to the total deposit of the payment hub.
Each participant then replies to the leader with its signatures on the epoch state $\textit{State}(e+1)$.
Once all signatures are collected by the leader, an epoch state confirmation containing both $\textit{State}(e+1)$ and all signatures will be broadcasted to all participants to finish the current epoch.
After that, participants of the payment hub can proceed to the next epoch.

\subsubsection{Dynamic Enrollment and Withdrawal} \label{dynamicEnrollmentAndWithdrawal}

Garou allows its participants to join in and withdraw from the payment hub at any time.
Each participant can join in multiple Garou instances at the same time.
To join an existing payment hub, the new participant should first invoke the on-chain smart contract instance along with its initial deposit.
As mentioned above, the on-chain transactions will be available for all participants, including the leader.
Then in the next consensus phase, the leader should add the new participant into the enrollment set $\mathcal{E}(e)$ and send it to all participants along with the new epoch state, as shown in equation (\ref{eqn:state}).
The enrollment set $\mathcal{E}(e)$ contains those newly joined participants and their initial deposit on the payment hub.
After reaching a consensus, the newly joined participant can then perform off-chain transactions with each other in the next epoch.

Similar to the joining process, we use the withdrawal set $\mathcal{W}(e)$ as an off-chain synchronization tool.
To withdraw from a payment hub, the participant should first send a withdrawal request to the leader.
When the leader receives that withdrawal request, he adds the participant into the withdrawal set and broadcasts it to all participants along with the new epoch state.
The withdrawal set $\mathcal{W}(e)$ contains those participants that are intended to withdraw and their final balance of epoch $e$.
After reaching a consensus, the participant intended to withdraw from the payment hub is not allowed to make any coin transfers with others.
The participant can now safely invoke the on-chain smart contract to reclaim its balance.
It should first commit an on-chain withdrawal initialization request along with the latest epoch number and its final balance.
After waiting for a predetermined time, e.g. $2T$, the participant commits an on-chain withdrawal confirmation transaction to retrieve its money.

\section{Security Analysis} \label{securityAnalysis}

Garou guarantees that an honest participant will not bear any financial losses despite strong adversarial capabilities, i.e., even when all other participants of the payment hub are malicious.
The balance security of honest participants is mainly enforced by the on-chain smart contract.
We firstly describe the threat model and then analyze the security properties of the Garou protocol.

\subsection{Threat Model} \label{threadModel}

Similar to other off-chain protocols, we assume the presence of an irrational adversary that can behave arbitrarily at any given time.
The irrational adversarial may sustain financial losses in order to cause honest participants to lose some or all of their funds.
Besides, it may take control of the leader, some or even all but one honest participant.
The internal states, communication channels, and identities of the corrupted participants are fully exposed to the adversary.
In addition, the adversary may send arbitrary messages on behalf of the corrupted party, launch a denial of service attack, etc.
We assume that the communication channels among honest participants and their integrity of identities cannot be corrupted by the adversary.
We also assume that the communication with the blockchain network are uncorrupted because nodes can participate in the blockchain network through any miner.
Furthermore, we assume that the adversary is computationally bounded, i.e., corrupting cryptographic primitives is unfeasible for an adversary.

\subsection{On-Chain Dispute Resolver} \label{onChainDisputeResolver}
As mentioned earlier, the on-chain smart contract is responsible for managing the deposits of all participants and acts as a dispute resolver for the payment hub.
We now discuss how the on-chain smart contract resolves disputes.

Garou allows any participant to open an on-chain epoch state challenge as long as it fails to receive the epoch state confirmation of the latest epoch, which may be caused by uncooperative participants or a malicious leader that intentionally withholds the latest epoch state from honest participants.
The epoch state challenge enforces the latest epoch state agreed by all participants of the payment hub.
The epoch state with the largest epoch id $e$ is exactly the latest one.
When a participant does not receive the epoch state confirmation upon timeout during the consensus phase, it will invoke the on-chain smart contract to open up an epoch state challenge.
The answer to that state challenge, which consists of the latest epoch state and signatures from all participants, should be committed to the contract by the leader within a predefined timeout $T$.
If the epoch state challenge is fulfilled, participants could proceed to the next epoch of off-chain execution.
Otherwise, the payment hub should roll back to the last epoch state that is agreed upon by all its participants.
In this case, participants can choose to proceed to the next epoch of the enforced one, or withdraw from the payment hub.

\subsection{Liveness} \label{liveness}

Since in each epoch all participants should confirm the epoch state $\textit{State}(e+1)$, Garou requires that all participants of the payment hub should always keep online.
A malicious participant could launch denial-of-service attacks to hinder the proceeding of the Garou protocol.
More specifically, a malicious node could (1) refuse to reply its signature on the epoch state $\textit{State}(e+1)$ to the leader in the consensus phase.
A malicious leader could (2) refuse to construct and broadcast the epoch state $\textit{State}(e+1)$ in the consensus phase, or (3) refuse to approve the transaction id in the trading phase.
All the above cases will block the proceeding of the Garou protocol.

For the case (3), it's quite trivial since one could wait until the next epoch to reapply for a transaction id.
The automatic leader election mechanism can reduce the risk of a malicious leader and prevent the malicious leader from dominating the payment hub for a long time.
For the case (2), since only the leader maintains all the transaction information of the current epoch, it's infeasible to construct the correct epoch state $\textit{State}(e+1)$ in the presence of a malicious leader.
In this case, other participants should open up an epoch state challenge to roll back to the previous state.
For the case (1), the epoch state $\textit{State}(e+1)$ is correctly constructed but remains incompleted due to a lack of signatures of offline participants.
In this case, the leader should commit the incompleted epoch state $\textit{State}^\prime(e+1)$ to the on-chain smart contract to enforce signatures.
When an incompleted epoch state is submitted to the on-chain contract, participants that have not yet provided its signature need to submit their signature on $\textit{State}^\prime(e+1)$ to the on-chain contract within a maximum timeout $T$.
Otherwise, those uncooperative participants will be evicted out by the on-chain dispute resolver.

\subsection{Security Properties} \label{securityProperties}

\subsubsection{Balance Security} \label{balanceSecurity}

Garou guarantees that an honest participant will not bear any financial losses even when all other participants are malicious.
Since the adversary cannot corrupt the identity of the honest participant, it is unfeasible to fabricate spending transactions in the name of honest participants.
Therefore the only way to cause a financial loss on an honest participant is to tamper the balance on the incompleted epoch state.
However, an honest participant always maintains knowledge about all its involved transactions and therefore is able to determine its final balance of the current epoch.
When receiving a new epoch state from the leader in the consensus phase, an honest participant should verify the right amount of its balance and the right amount of total balance of the payment hub.
Otherwise, an honest participant should refuse to reply with its signature on the epoch state that may lead to its financial loss.
This will effectively prevent the payment hub from reaching a consensus, and the result is an on-chain epoch state challenge, which enforces the leader to provide a properly constructed epoch state.

\subsubsection{Transaction Confirmation} \label{transactionConfirmation}
Garou guarantees that if the leader is honest, all valid transactions of an honest participant are expected to be confirmed at the end of the current epoch.
However, in the presence of a malicious leader, there is no guarantee that transactions will eventually be confirmed.
A valid transaction consists of a transaction id signed by the leader, the right amount of coins, and signatures from both the sender and receiver.
When constructing and verifying the transaction merkle tree root $\mathcal{M}(e)$, one should always ensure that all of its involved transactions, including both spending and receiving, are included and correctly ordered by the transaction id.
Given an ordered set of transactions as its leaves, the constructed merkle tree root will be unique.
In this way, a properly constructed merkle tree can ensure that all valid transactions of an honest participant will be eventually confirmed.
However, a malicious leader may choose to omit some of the valid transactions when constructing the merkle tree root.
In this case, an honest participant should refuse to reply with its signature on the epoch state and resort to the on-chain smart contract to enforce a properly constructed epoch state.

In fact, the confirmation time of an off-chain transaction is exactly equal to the epoch duration, which is the same as NOCUST.
And for this reason, Garou does not allow those unconfirmed coins to be spent until the next epoch.
However, NOCUST does allow its coins to be spent before final confirmation, which increases the overall risk and therefore demands a shorter epoch duration.
Although the shortening of epoch duration allows transactions to be confirmed in a timely manner, it also comes with the cost of overall transaction throughput.
Users of Garou would have to make the trade-off between the confirmation time and transaction throughput.
Our subsequent evaluation suggests an epoch duration of 10 seconds.

\subsubsection{Double Spending} \label{doubleSpending}
The support for concurrent transactions will lead to a situation where a participant cannot accurately know the available balance of other participants.
Therefore a malicious participant could have a chance to launch a double-spending attack by spending more than its available balance.
The problem is quite trivial if the leader is honest.
When issuing the transaction id, the leader should verify that a participant $\mathcal{P}_i$ should not spend more than its initial balance $\mathcal{B}_i(e)$ in the current epoch.
If someone is trying to over-spend its balance, an honest leader should refuse to reply with transaction id, which effectively prevents the malicious participant from double-spending.
However, if the leader happens to be malicious, the situation becomes more complicated.
In this case, a malicious participant will manage to get a transaction id from the leader and construct a spending transaction that overspends its balance to an honest receiver.
However, an over-spending transaction will eventually cause the total balance of the payment hub to deviate from its correct value $b$, more specifically, $\sum_{i=1}^{n}\mathcal{B}_i(e) > b$.
Therefore, an honest participant should verify the right amount of the payment hub and ensure that $\sum_{i=1}^{n}\mathcal{B}_i(e) \equiv b$ holds during the consensus phase.
Otherwise, it should refuse to provide its signature on the epoch state and open an on-chain state challenge to enforce its balance security.
In the worst case, the payment hub should roll back to the last epoch state that is agreed upon by all its participants.

\subsubsection{Privacy} \label{privacy}

Garou requires that the leader should maintain complete knowledge of all off-chain transactions to construct the new epoch state properly during the consensus phase.
Therefore Garou provides no privacy guarantees in the presence of a malicious leader, i.e., a malicious leader may leak all transaction information of the current epoch in the worst case.
We note that the information leakage is only limited to the current epoch because a malicious leader cannot obtain transaction information of the last epoch.
On the other hand, if the leader is honest, a malicious participant can only learn information about the initial balance of other participants, i.e. through $\mathcal{B}(e)$.
Apart from this, it cannot learn any information about the transaction history of other participants.

\section{Implementation} \label{implementation}

To evaluate the effectiveness and performance of our design, we implement the Garou protocol for the Ethereum network.
Then we measure the execution cost, evaluate the off-chain performance of the Garou protocol, and compare it with the state-of-art payment hubs.

\subsection{Execution Cost} \label{executionCost}

\begin{table}[!t]
    \renewcommand{\arraystretch}{1.3}
    \caption{Execution Cost and Message Complexity of Garou}
    \label{tbl:cost}
    \centering
    \begin{tabular}{c||c|c|c}
    \hline
    \bfseries Operation & \bfseries On-Chain Tx & \bfseries Gas Cost & \bfseries Off-Chain Msg \\
    \hline\hline
    Join                & 1                     & 43211              & 0                       \\
    \hline
    Withdraw & 2 & 100166
    & 0 \\
    \hline
    Transfer            & 0                     & 0                  & 5                       \\ 
    \hline
    \begin{tabular}{l}
    Consensus \\ \scriptsize{(optimistic)}
\end{tabular} & 0 & 0 & $3n$ \\
\hline
\begin{tabular}{l}
    Consensus \\ \scriptsize{(pessimistic)}
\end{tabular} & $1+m$ & $10962n$ & - \\
\hline
\multicolumn{4}{l}{n: the number of participants of payment hub} \\
\multicolumn{4}{l}{m: the number of malicous participants} \\
\end{tabular}
\end{table}

Ethereum uses gas to measure the amount of computational and storage resources cost by the execution of smart contracts.
Every instruction executed by the \textit{Ethereum Virtual Machine} (EVM) costs a certain amount of gas, e.g., verifying a user's signature costs 3000 gas units \cite{buterin2014ethereum}. 
Table \ref{tbl:cost} shows the execution costs and off-chain message complexity of each operation of the Garou protocol.
Apart from the joining and withdrawal process, if all participants are honest and cooperative, the off-chain coin transfers and consensus do not require any involvements of the blockchain, resulting in zero on-chain transaction and gas cost.
Every off-chain coin transfer costs a constant five messages, and the message complexity of the consensus phase is linear to the number of participants.
In the presence of malicious parties, no matter a malicious leader or ordinary participants, $1+m$ on-chain transactions are needed to resolve disputes:
one for opening an epoch state challenge, and $m$ for each malicious party answering the state challenge.
The gas cost of answering an epoch state challenge mainly consists of verifying participants' signatures on the epoch state, and the cost used to commit the epoch state data to the smart contract.

In Ethereum, the gas limit refers to the upper bound of gas that can be consumed by executing a smart contract.
As more participants join in the payment hub, more information and signatures will need to be collected and verified by the on-chain smart contract when resolving disputes.
Therefore, the number of participants of a payment hub cannot grow infinitely.
In our implementation, the cost of resolving disputes increases by approximately 10962 gas units per node (as shown in Table \ref{tbl:cost}), leading to the maximum number of participants to be approximately 300 for each payment hub.
However, as mentioned in Section \ref{blockchain}, the blockchain network allows multiple independent contract instances to be running at the same time.
Therefore we could scale out the off-chain system by instantiating multiple payment hubs.

\subsection{Performance Evalution} \label{performanceEvaluation}

We now evaluate the off-chain transaction performance of the Garou protocol and compare it with NOCUST.
Our subsequent evaluations are conducted on one single payment hub, i.e., we open a payment hub instance, let participants join in the payment hub, and then measure the off-chain transaction throughput and latency of that payment hub instance.
Our evaluations are run on a machine with Intel(R) Core(TM) i5-8400 2.80GHz CPU and 32G RAM.
The network bandwidth between each participant is set to 20Mbps and the average latency is about $100 \pm 20$ ms.

To effectively measure the maximum off-chain performance, we need to assume that all participants of the payment hub are honest and cooperative.
Otherwise, in the presence of malicious or unresponsive participants, the honest one should pause the off-chain processes and invoke the on-chain smart contract, which will lead to an unpredictable delay, depending on the behavior of malicious participants and the delay of the blockchain networks.
Since there is no realistic off-chain transaction dataset, and the available on-chain transaction datasets cannot meet our concurrency requirement.
Therefore we use randomly generated transaction dataset in our evaluation.

\subsubsection{Transaction Throughput} \label{transactionThroughput}

\begin{figure}[!t]
    \centering
    \begin{subfigure}[b]{0.4\textwidth}
        \centering
        \begin{tikzpicture}
	\begin{axis}[
			axis x line*=bottom,
			axis y line*=left,
			axis z line*=left,
			nodes near coords,
			every node near coord/.append style={font=\footnotesize, /pgf/number format/1000 sep={}},
			grid=major,
			x dir=reverse,
			xmin=5,xmax=55,
			ymin=1,ymax=9,
			xtick={10, 20, 30, 40, 50},
			xticklabels={10,50,100,200,300},
			ytick={2,4,6,8},
			yticklabels={2,5,10,60},
			ztick={6000, 7000, 8000,9000},
			z tick label style={/pgf/number format/1000 sep={}},
			xlabel=Number of Nodes,
			ylabel=Epoch (s),
			zlabel=Throughput (tps),
			xlabel style={sloped like x axis, yshift=1ex},
			ylabel style={sloped like y axis, yshift=1ex},
			zlabel style={rotate=180, xshift=-2ex, yshift=1ex}
		]
		
		\addplot3+[
			surf,
			scatter,
		mesh/rows=5]
		coordinates {
			(10, 2, 5514)
			(10, 4, 7676)
			(10, 6, 8460)
			(10, 8, 8947)
			(20, 2, 5358)
			(20, 4, 7552)
			(20, 6, 8400)
			(20, 8, 9064)
			(30, 2, 5470)
			(30, 4, 7642)
			(30, 6, 8474)
			(30, 8, 9153)
			(40, 2, 5626)
			(40, 4, 7861)
			(40, 6, 8619)
			(40, 8, 9299)
			(50, 2, 5439)
			(50, 4, 7782)
			(50, 6, 8714)
			(50, 8, 9412)
		};
	\end{axis}
\end{tikzpicture}
        \caption{Garou}
        \label{fig:garouTPS}
    \end{subfigure}
    \hfill
    \begin{subfigure}[b]{0.4\textwidth}
        \centering
        \begin{tikzpicture}
	\begin{axis}[
			axis x line*=bottom,
			axis y line*=left,
			axis z line*=left,
			nodes near coords,
			every node near coord/.append style={font=\footnotesize, /pgf/number format/1000 sep={}},
			grid=major,
			x dir=reverse,
			xmin=5,xmax=55,
			ymin=1,ymax=9,
			xtick={10, 20, 30, 40, 50},
			xticklabels={10,50,100,200,300},
			ytick={2,4,6,8},
			yticklabels={2,5,10,60},
			ztick={100,200,300,400},
			z tick label style={/pgf/number format/1000 sep={}},
			xlabel=Number of Nodes,
			ylabel=Epoch (s),
			zlabel=Throughput (tps),
			xlabel style={sloped like x axis, yshift=1ex},
			ylabel style={sloped like y axis, yshift=1ex},
			zlabel style={rotate=180, xshift=-2ex, yshift=1ex}
		]
			
		\addplot3+[
			surf,
			scatter,
		mesh/rows=5]
		coordinates {
			(10,2,14)
			(10,4,15)
			(10,6,16)
			(10,8,16)
			(20,2,70)
			(20,4,78)
			(20,6,80)
			(20,8,82)
			(30,2,140)
			(30,4,155)
			(30,6,159)
			(30,8,164)
			(40,2,282)
			(40,4,310)
			(40,6,316)
			(40,8,328)
			(50,2,422)
			(50,4,465)
			(50,6,475)
			(50,8,492)
		};
	\end{axis}
\end{tikzpicture}
        \caption{NOCUST}
        \label{fig:nocustTPS}
    \end{subfigure}
    \caption{Impact of different number of nodes and epoch duration on maximum off-chain transaction throughput}
    \label{fig:throughput}
\end{figure}
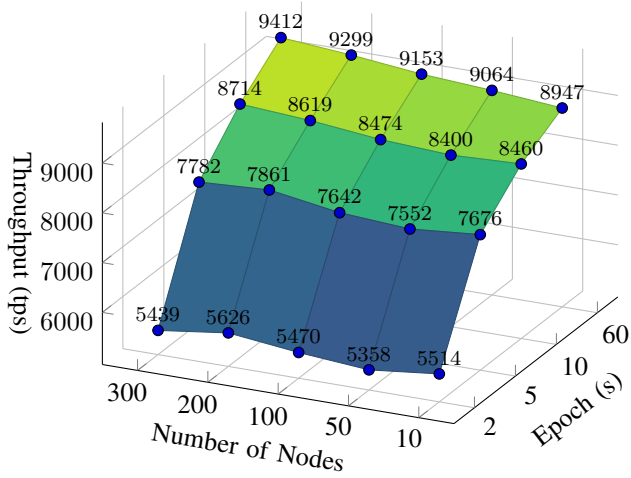
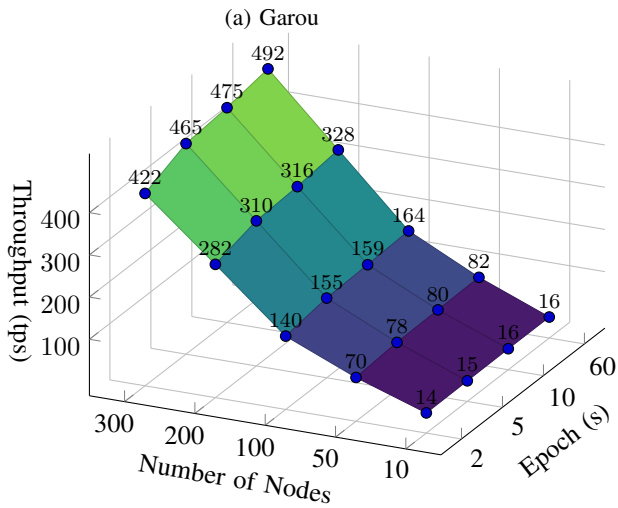

Firstly, we measure the maximum transaction throughput under different combinations of the number of nodes and epoch duration.
Specifically, we vary the number of nodes from 10 to 300.
Besides, the epoch duration is closely related to the confirmation time of an off-chain transaction for both Garou and NOCUST.
We do not expect a long confirmation time and therefore vary the epoch duration from 2 seconds to 60 seconds.
Fig. \ref{fig:throughput} summaries the maximum off-chain transaction throughput of Garou and NOCUST respectively.
We do not measure the transaction throughput of the multi-party virtual state channel \cite{dziembowski2019multi} since its upper limit is quite predictable.
It requires that any state update should be explicitly agreed upon by all participants.
Therefore, given that the network latency is about $100$ ms, its theoretical transaction throughput cannot exceed 10. 

From the results of our evaluation, we notice that the off-chain transaction throughput of Garou is much higher than NOCUST.
Garou obtains quite high transaction throughput when the epoch duration is set to 10 seconds.
In this case, Garou exhibits nearly $20 \times$ the maximum throughput of NOCUST.
However, the transaction throughput of these two payment hubs demonstrates different growth trends.
Garou's transaction throughput increases with the length of the epoch duration and does not increase with the number of participants, but it is just the opposite for NOCUST.
This can be explained as below.

Garou's transaction throughput is mainly governed by the leader because every off-chain coin transfer requires a transaction id issued by the leader.
Therefore, the increase in the number of participants will not lead to an increase in transaction throughput.
On the contrary, the shortening of epoch duration will cause the leader to spend a larger proportion of time in the consensus phase, and therefore reduces the overall transaction throughput of the payment hub.
This is quite obvious when the epoch duration is relatively short, e.g., 2s.
On the contrary, NOCUST does not allow both the sender and receiver to engage in any other transfers until the current one is settled, which significantly limits the concurrency and the transaction throughput of the payment hub.
As a result, the maximum transaction throughput of NOCUST is much worse than Garou.
The only way to increase the concurrency of NOCUST, as shown in Fig. \ref{fig:nocustTPS}, is letting more pair of nodes to make transfers simultaneously. However, as mentioned above, the number of participants cannot be increased infinitely due to the gas limit.

As shown in Fig.\ref{fig:garouTPS}, the off-chain transaction throughput of Garou protocol increases significantly when the epoch duration is between 2 and 10 seconds. 
When the epoch duration exceeds 10 seconds, the growth of transaction throughput slows down.
Moreover, a longer epoch duration means longer transaction confirmation time.
Therefore, to achieve a balance between transaction throughput and confirmation time, we set the epoch duration to 10 seconds in the following evaluations.

\subsubsection{Transaction Latency} \label{transactionLatency}

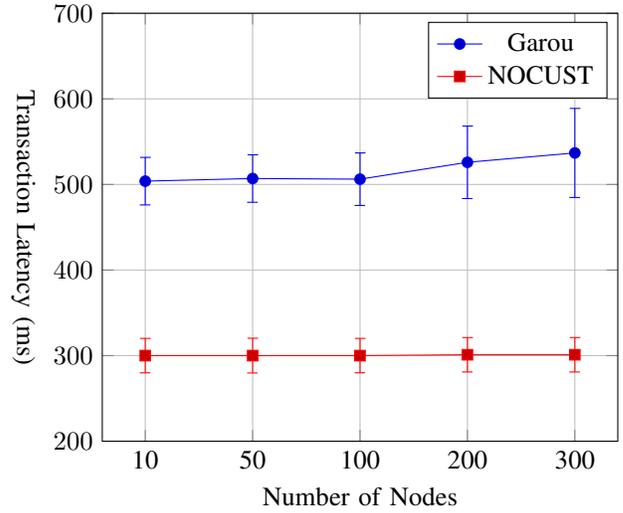
\begin{figure}[!t]
    \centering
    \begin{tikzpicture}
    \begin{axis}[
        grid=major,
        xtick={10, 20, 30, 40, 50},
        xticklabels={10,50,100,200,300},
        xlabel=Number of Nodes,
        ymin=200,ymax=700,
        ylabel=Transaction Latency (ms),
        ylabel style={rotate=180},
    ]
    \addplot+[
        error bars/.cd,
        y dir=both,
        y explicit,
    ]
    coordinates {
        (10, 503.8389045)+-(0, 27.76876543)
        (20, 506.9241663)+-(0, 27.74175097)
        (30, 506.148837)+-(0, 30.68594136)
        (40, 525.8652125)+-(0, 42.34456146)
        (50, 536.8004375)+-(0, 52.03654111)
    }; \addlegendentry{Garou}
    \addplot+[
        error bars/.cd,
        y dir=both,
        y explicit,
    ]
    coordinates {
        (10, 300) +- (0, 20.11071226)
        (20, 300) +- (0, 20.29402075)
        (30, 300) +- (0, 20.08349504)
        (40, 301) +- (0, 20.09905889)
        (50, 301) +- (0, 20.11812672)
    }; \addlegendentry{NOCUST}
    \end{axis}
\end{tikzpicture}
    \caption{Average Transaction Latency}
    \label{fig:transactionLatency}
\end{figure}

Fig. \ref{fig:transactionLatency} shows the average off-chain transaction latency of both Garou and NOCUST.
As expected, the concurrency limitation of NOCUST leads to its smooth latency curve.
In contrast, the transaction latency of Garou remains stable when the number of participants does not exceed 100.
When it does exceed 100, the transaction latency starts to increase gradually.
Nevertheless, even when the number of participants reaches the upper limit of 300, the average transaction latency still does not exceed 1 second, which is quite acceptable.
We also notice that the average transaction latency of Garou is higher than NOCUST.
This is because Garou's off-chain transactions require more communication between the leader and participants to prevent malicious from double-spending in concurrent scenarios.
At the cost in transaction latency, Garou obtains an order of magnitude improvement in throughput.

\subsubsection{Consensus Delay} \label{consensusDelay}

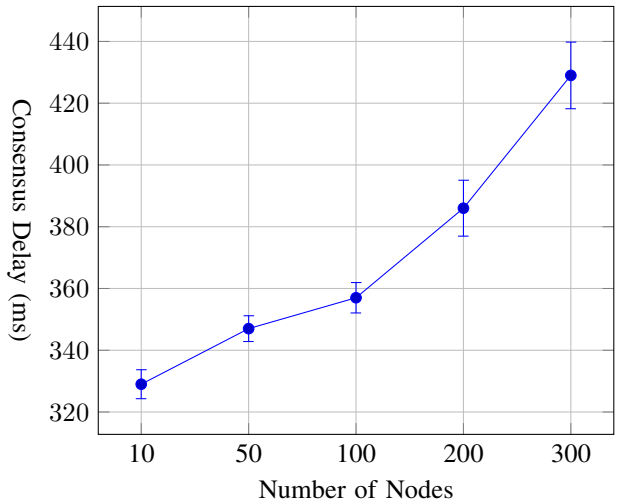
\begin{figure}[!t]
    \centering
    \begin{tikzpicture}
    \begin{axis}[
        grid=major,
        xtick={10, 20, 30, 40, 50},
        xlabel=Number of Nodes,
        xticklabels={10,50,100,200,300},
        ylabel=Consensus Delay (ms),
        ylabel style={rotate=180},
    ]
        \addplot+[
            error bars/.cd,
            y dir=both,
            y explicit,
        ]
        coordinates {
            (10, 329) +- (0, 4.69)
            (20, 347) +- (0, 4.17)
            (30, 357) +- (0, 4.92)
            (40, 386) +- (0, 9.05)
            (50, 429) +- (0, 10.79)
        };
    \end{axis}
\end{tikzpicture}
    \caption{Consensus Delay of Garou}
    \label{fig:consensusDelay}
\end{figure}

Fig. \ref{fig:consensusDelay} shows the consensus delay of Garou protocol, i.e., the time spent in the consensus phase for each epoch.
Since NOCUST does not have a consensus phase, we only test the consensus delay of our proposed protocol.
During the consensus phase, all participants of the payment hub are not allowed to make any coin transfers until entering the next epoch.
Similar to transaction latency, the consensus delay starts to increase significantly when the number of participants exceeds 100.
This is simply because the leader needs to communicate with all participants of the payment hub and collect their signatures in the consensus phase.

\subsubsection{Summary of Results} \label{summaryOfResults}

In summary, the performance evaluations show that Garou has achieved a substantial improvement in transaction throughput than the state-of-art payment hubs.
More precisely, the maximum transaction throughput of Garou is $20 \times$ higher than NOCUST.
Roughly, Garou's high performance mainly benefits from its support for concurrent off-chain transactions.
Moreover, when the number of participants is within 100 and the epoch duration is set to 10 seconds, Garou can achieve a balance in both transaction throughput, latency, and consensus delay.
Since there is no limitation on the number of Garou instances, our proposed protocol could dramatically improve the performance and scalability of the blockchain systems.


\section{Conclusion} \label{conclusion}

Currently, the off-blockchain payment system is still far from maturity and exhibits many challenges in terms of performance, scalability, privacy, and security.
In this paper, we present Garou, a high-performance and secure off-chain N-party payment hub that allows multiple parties to perform concurrent and flexible off-chain coin transfers with each other without compromising security.
The support for concurrent transaction significantly increases the performance of the payment hub and allows Garou to achieve up to $20 \times$ the maximum transaction throughput of the state-of-art payment hubs.
Transactions within the payment hub can be executed directly between participants without the need of any third-party operators.
Garou also guarantees the balance security of honest participants despite strong adversarial capabilities.
Overall, our proposed Garou protocol achieves a balance in both performance, flexibility, and security and therefore serves as an enrichment to the off-blockchain payment systems.



%


\bibliographystyle{IEEEtran}
\bibliography{references}

\begin{thebibliography}{10}
\providecommand{\url}[1]{#1}
\csname url@samestyle\endcsname
\providecommand{\newblock}{\relax}
\providecommand{\bibinfo}[2]{#2}
\providecommand{\BIBentrySTDinterwordspacing}{\spaceskip=0pt\relax}
\providecommand{\BIBentryALTinterwordstretchfactor}{4}
\providecommand{\BIBentryALTinterwordspacing}{\spaceskip=\fontdimen2\font plus
\BIBentryALTinterwordstretchfactor\fontdimen3\font minus
  \fontdimen4\font\relax}
\providecommand{\BIBforeignlanguage}[2]{{%
\expandafter\ifx\csname l@#1\endcsname\relax
\typeout{** WARNING: IEEEtran.bst: No hyphenation pattern has been}%
\typeout{** loaded for the language `#1'. Using the pattern for}%
\typeout{** the default language instead.}%
\else
\language=\csname l@#1\endcsname
\fi
#2}}
\providecommand{\BIBdecl}{\relax}
\BIBdecl

\bibitem{nakamoto2008bitcoin}
S.~Nakamoto \emph{et~al.}, ``Bitcoin: A peer-to-peer electronic cash system,''
  2008.

\bibitem{dwork1992pricing}
C.~{Dwork} and M.~{Naor}, ``Pricing via processing or combatting junk mail,''
  in \emph{CRYPTO '92 Proceedings of the 12th Annual International Cryptology
  Conference on Advances in Cryptology}, 1992, pp. 139--147.

\bibitem{king2012ppcoin}
S.~King and S.~Nadal, ``Ppcoin: Peer-to-peer crypto-currency with
  proof-of-stake,'' \emph{self-published paper, August}, vol.~19, 2012.

\bibitem{wood2014ethereum}
G.~Wood, ``Ethereum: A secure decentralised generalised transaction ledger,''
  \emph{Ethereum project yellow paper}, vol. 151, pp. 1--32, 2014.

\bibitem{croman2016scaling}
K.~Croman, C.~Decker, I.~Eyal, A.~E. Gencer, A.~Juels, A.~Kosba, A.~Miller,
  P.~Saxena, E.~Shi, E.~G. Sirer \emph{et~al.}, ``On scaling decentralized
  blockchains,'' in \emph{International Conference on Financial Cryptography
  and Data Security}.\hskip 1em plus 0.5em minus 0.4em\relax Springer, 2016,
  pp. 106--125.

\bibitem{gervais2016security}
A.~Gervais, G.~O. Karame, K.~W{\"u}st, V.~Glykantzis, H.~Ritzdorf, and
  S.~Capkun, ``On the security and performance of proof of work blockchains,''
  in \emph{Proceedings of the 2016 ACM SIGSAC Conference on Computer and
  Communications Security}.\hskip 1em plus 0.5em minus 0.4em\relax ACM, 2016,
  pp. 3--16.

\bibitem{trillo2013stress}
\BIBentryALTinterwordspacing
M.~Trillo, ``Stress test prepares visanet for the most wonderful time of the
  year (2013),'' 2013. [Online]. Available:
  \url{https://www.visa.com/blogarchives/us/2013/10/10/stress-test-prepares-visanet-for-the-most-wonderful-time-of-the-year/index.html}
\BIBentrySTDinterwordspacing

\bibitem{decker2015fast}
C.~{Decker} and R.~{Wattenhofer}, ``A fast and scalable payment network with
  bitcoin duplex micropayment channels,'' in \emph{Proceedings of the 17th
  International Symposium on Stabilization, Safety, and Security of Distributed
  Systems - Volume 9212}, 2015, pp. 3--18.

\bibitem{poon2016bitcoin}
J.~Poon and T.~Dryja, ``The bitcoin lightning network: Scalable off-chain
  instant payments,'' \emph{See https://lightning.
  network/lightning-network-paper. pdf}, 2016.

\bibitem{malavolta2017concurrency}
G.~{Malavolta}, P.~{Moreno-Sanchez}, A.~{Kate}, M.~{Maffei}, and S.~{Ravi},
  ``Concurrency and privacy with payment-channel networks,'' in
  \emph{Proceedings of the 2017 ACM SIGSAC Conference on Computer and
  Communications Security}, vol. 2017, 2017, pp. 455--471.

\bibitem{ethan2017tumblebit}
H.~Ethan, F.~Baldimtsi, L.~Alshenibr, A.~Scafuro, and S.~Goldberg, ``Tumblebit:
  An untrusted tumbler for bitcoin-compatible anonymous payments,'' in
  \emph{Network and Distributed System Security Symposium (NDSS)}, 2017.

\bibitem{dziembowski2019multi}
S.~Dziembowski, L.~Eckey, S.~Faust, J.~Hesse, and K.~Host{\'a}kov{\'a},
  ``Multi-party virtual state channels,'' in \emph{Annual International
  Conference on the Theory and Applications of Cryptographic Techniques}.\hskip
  1em plus 0.5em minus 0.4em\relax Springer, 2019, pp. 625--656.

\bibitem{khalil2018nocust}
R.~Khalil and A.~Gervais, ``Nocust-a non-custodial 2nd-layer financial
  intermediary.'' \emph{IACR Cryptology ePrint Archive}, vol. 2018, p. 642,
  2018.

\bibitem{eyal2016bitcoin}
I.~{Eyal}, A.~E. {Gencer}, E.~G. {Sirer}, and R.~van {Renesse}, ``Bitcoin-ng: a
  scalable blockchain protocol,'' \emph{networked systems design and
  implementation}, pp. 45--59, 2016.

\bibitem{luu2015scp}
L.~{Luu}, V.~{Narayanan}, K.~{Baweja}, C.~{Zheng}, S.~{Gilbert}, and
  P.~{Saxena}, ``Scp: A computationally-scalable byzantine consensus protocol
  for blockchains.'' \emph{IACR Cryptology ePrint Archive}, vol. 2015, p. 1168,
  2015.

\bibitem{pass2017hybrid}
R.~{Pass} and E.~{Shi}, ``Hybrid consensus: Efficient consensus in the
  permissionless model,'' \emph{international conference on distributed
  computing}, vol.~91, p.~16, 2017.

\bibitem{lind2017teechain}
J.~{Lind}, I.~{Eyal}, F.~{Kelbert}, O.~{Naor}, P.~R. {Pietzuch}, and E.~G.
  {Sirer}, ``Teechain: Scalable blockchain payments using trusted execution
  environments.'' 2017.

\bibitem{zhang2016town}
F.~{Zhang}, E.~{Cecchetti}, K.~{Croman}, A.~{Juels}, and E.~{Shi}, ``Town
  crier: An authenticated data feed for smart contracts,'' in \emph{Proceedings
  of the 2016 ACM SIGSAC Conference on Computer and Communications Security},
  vol. 2016, 2016, pp. 270--282.

\bibitem{zamani2018rapidchain}
M.~{Zamani}, M.~{Movahedi}, and M.~{Raykova}, ``Rapidchain: Scaling blockchain
  via full sharding,'' in \emph{CCS '18 Proceedings of the 2018 ACM SIGSAC
  Conference on Computer and Communications Security}, 2018, pp. 931--948.

\bibitem{luu2016secure}
L.~{Luu}, V.~{Narayanan}, C.~{Zheng}, K.~{Baweja}, S.~{Gilbert}, and
  P.~{Saxena}, ``A secure sharding protocol for open blockchains,'' in
  \emph{Proceedings of the 2016 ACM SIGSAC Conference on Computer and
  Communications Security}, 2016, pp. 17--30.

\bibitem{kokoris-kogias2018omniledger}
E.~{Kokoris-Kogias}, P.~{Jovanovic}, L.~{Gasser}, N.~{Gailly}, E.~{Syta}, and
  B.~{Ford}, ``Omniledger: A secure, scale-out, decentralized ledger via
  sharding,'' in \emph{2018 IEEE Symposium on Security and Privacy (SP)}, 2018,
  pp. 583--598.

\bibitem{nguyen2019optchain}
L.~N. Nguyen, T.~D. Nguyen, T.~N. Dinh, and M.~T. Thai, ``Optchain: Optimal
  transactions placement for scalable blockchain sharding,'' in \emph{2019 IEEE
  39th International Conference on Distributed Computing Systems
  (ICDCS)}.\hskip 1em plus 0.5em minus 0.4em\relax IEEE, 2019, pp. 525--535.

\bibitem{mccorry2016towards}
P.~McCorry, M.~M{\"o}ser, S.~F. Shahandasti, and F.~Hao, ``Towards bitcoin
  payment networks,'' in \emph{Australasian Conference on Information Security
  and Privacy}.\hskip 1em plus 0.5em minus 0.4em\relax Springer, 2016, pp.
  57--76.

\bibitem{wang2019monoxide}
J.~Wang and H.~Wang, ``Monoxide: Scale out blockchains with asynchronous
  consensus zones,'' in \emph{16th $\{$USENIX$\}$ Symposium on Networked
  Systems Design and Implementation ($\{$NSDI$\}$ 19)}, 2019, pp. 95--112.

\bibitem{chauhan2018blockchain}
A.~Chauhan, O.~P. Malviya, M.~Verma, and T.~S. Mor, ``Blockchain and
  scalability,'' in \emph{2018 IEEE International Conference on Software
  Quality, Reliability and Security Companion (QRS-C)}.\hskip 1em plus 0.5em
  minus 0.4em\relax IEEE, 2018, pp. 122--128.

\bibitem{decker2018eltoo}
C.~Decker, R.~Russell, and O.~Osuntokun, ``eltoo: A simple layer2 protocol for
  bitcoin,'' \emph{White paper: https://blockstream. com/eltoo. pdf}, 2018.

\bibitem{green2017bolt}
M.~D. {Green} and I.~{Miers}, ``Bolt: Anonymous payment channels for
  decentralized currencies,'' in \emph{Proceedings of the 2017 ACM SIGSAC
  Conference on Computer and Communications Security}, vol. 2016, 2017, pp.
  473--489.

\bibitem{hopwood2016zcash}
D.~Hopwood, S.~Bowe, T.~Hornby, and N.~Wilcox, ``Zcash protocol
  specification,'' \emph{Tech. rep. 2016--1.10. Zerocoin Electric Coin Company,
  Tech. Rep.}, 2016.

\bibitem{miers2013zerocoin}
I.~Miers, C.~Garman, M.~Green, and A.~D. Rubin, ``Zerocoin: Anonymous
  distributed e-cash from bitcoin,'' in \emph{2013 IEEE Symposium on Security
  and Privacy}.\hskip 1em plus 0.5em minus 0.4em\relax IEEE, 2013, pp.
  397--411.

\bibitem{dziembowski2019perun}
S.~Dziembowski, L.~Eckey, S.~Faust, and D.~Malinowski, ``Perun: Virtual payment
  hubs over cryptocurrencies,'' in \emph{2019 IEEE Symposium on Security and
  Privacy (SP)}.\hskip 1em plus 0.5em minus 0.4em\relax IEEE, 2019, pp.
  106--123.

\bibitem{dziembowski2018general}
\BIBentryALTinterwordspacing
S.~Dziembowski, S.~Faust, and K.~Host\'{a}kov\'{a}, ``General state channel
  networks,'' in \emph{Proceedings of the 2018 ACM SIGSAC Conference on
  Computer and Communications Security}, ser. CCS '18.\hskip 1em plus 0.5em
  minus 0.4em\relax New York, NY, USA: ACM, 2018, pp. 949--966. [Online].
  Available: \url{http://doi.acm.org/10.1145/3243734.3243856}
\BIBentrySTDinterwordspacing

\bibitem{malavolta2019anonymous}
G.~Malavolta, P.~Moreno-Sanchez, C.~Schneidewind, A.~Kate, and M.~Maffei,
  ``Anonymous multi-hop locks for blockchain scalability and
  interoperability.'' in \emph{NDSS}, 2019.

\bibitem{prihodko2016flare}
P.~Prihodko, S.~Zhigulin, M.~Sahno, A.~Ostrovskiy, and O.~Osuntokun, ``Flare:
  An approach to routing in lightning network,'' \emph{White Paper (bitfury.
  com/content/5-white-papers-research/whitepaper\_flare\_an\_approach\_to\_routing\_in\_lightning\_n
  etwork\_7\_7\_2016. pdf)}, 2016.

\bibitem{malavolta2017silentwhispers}
G.~{Malavolta}, P.~{Moreno-Sanchez}, A.~{Kate}, and M.~{Maffei},
  ``Silentwhispers: Enforcing security and privacy in decentralized credit
  networks.'' in \emph{Network and Distributed System Security Symposium}, vol.
  2016, 2017, p. 1054.

\bibitem{roos2018settling}
S.~{Roos}, P.~{Moreno-Sanchez}, A.~{Kate}, and I.~{Goldberg}, ``Settling
  payments fast and private: Efficient decentralized routing for path-based
  transactions.'' in \emph{Proceedings 2018 Network and Distributed System
  Security Symposium}, 2018.

\bibitem{buterin2014ethereum}
V.~Buterin \emph{et~al.}, ``Ethereum: A next-generation smart contract and
  decentralized application platform,'' \emph{URL https://github.
  com/ethereum/wiki/wiki/\% 5BEnglish\% 5D-White-Paper}, vol.~7, 2014.

\bibitem{merkle1987digital}
R.~C. Merkle, ``A digital signature based on a conventional encryption
  function,'' in \emph{Conference on the theory and application of
  cryptographic techniques}.\hskip 1em plus 0.5em minus 0.4em\relax Springer,
  1987, pp. 369--378.

\end{thebibliography}

\end{document}